\documentclass[10pt,letterpaper]{article}
\usepackage[top=0.85in,left=2.75in,footskip=0.75in,marginparwidth=2in]{geometry}
\usepackage{booktabs}

\usepackage{amsmath,amsfonts,amssymb}

\usepackage{float}
\usepackage{url}
\usepackage{bbold}
\usepackage{multirow,tabularx}

\usepackage[utf8]{inputenc}

\usepackage{cite}

\usepackage{nameref,hyperref}

\usepackage[right]{lineno}

\usepackage{microtype}
\DisableLigatures[f]{encoding = *, family = * }

\raggedright
\usepackage{changepage}

\usepackage[aboveskip=1pt,labelfont=bf,labelsep=period,singlelinecheck=off]{caption}

\makeatletter
\renewcommand{\@biblabel}[1]{\quad#1.}
\makeatother

\usepackage{lastpage,fancyhdr,graphicx}
\usepackage{epstopdf}
\fancyheadoffset[L]{2.25in}
\fancyfootoffset[L]{2.25in}

\usepackage{color}

\definecolor{Gray}{gray}{.25}

\usepackage{graphicx}

\usepackage{sidecap}

\usepackage{wrapfig}
\usepackage[pscoord]{eso-pic}
\usepackage[fulladjust]{marginnote}
\reversemarginpar

\begin{document}
\vspace*{0.35in}

\begin{flushleft}
{\Large
\textbf\newline{Neuro-symbolic representation learning on biological knowledge graphs}
}
\newline
\\
Mona Alshahrani\textsuperscript{1},
Mohammed Asif Khan\textsuperscript{1},
Omar Maddouri\textsuperscript{1,2},
Akira R Kinjo\textsuperscript{3},
N\'uria Queralt-Rosinach\textsuperscript{4},
Robert Hoehndorf\textsuperscript{1,*}
\\
\bigskip

\bf{1} Computer, Electrical and Mathematical
Sciences \& Engineering Division, Computational Bioscience Research
Center, King Abdullah University of Science and Technology, 4700
KAUST, Thuwal 23955-6900, Kingdom of Saudi Arabia\\
\bf{2} Life Sciences Division, College of Science \& Engineering,
Hamad Bin Khalifa University, HBKU, PO Box 5825, Doha,
Qatar\\
\bf{3} Institute for Protein Research, Osaka University
3-2 Yamadaoka, Suita, Osaka, 565-0871, Japan\\
\bf{4} Department of Molecular and Experimental Medicine,
  The Scripps Research Institute, 10550 North Torrey Pines Road, La
  Jolla, CA 92037, USA.
\\
\bigskip
* robert.hoehndorf@kaust.edu.sa

\end{flushleft}

\section*{Abstract}
\textbf{Motivation:} Biological data and knowledge bases
  increasingly rely on Semantic Web technologies and the use of
  knowledge graphs for data integration, retrieval and federated
  queries. In the past years, feature learning methods that are
  applicable to graph-structured data are becoming available, but have
  not yet widely been applied and evaluated on structured biological
  knowledge.\\
  \textbf{Results:} We develop a novel method for feature learning on
  biological knowledge graphs. Our method combines symbolic methods,
  in particular knowledge representation using symbolic logic and
  automated reasoning, with neural networks to generate embeddings of
  nodes that encode for related information within knowledge
  graphs. Through the use of symbolic logic, these embeddings contain
  both explicit and implicit information. We apply these embeddings to
  the prediction of edges in the knowledge graph representing problems
  of function prediction, finding candidate genes of diseases,
  protein-protein interactions, or drug target relations, and
  demonstrate performance that matches and sometimes outperforms
  traditional approaches based on manually crafted features. Our
  method can be applied to any biological knowledge graph, and will
  thereby open up the increasing amount of Semantic Web based
  knowledge bases in biology to use in machine learning and data
  analytics. \\
  \textbf{Availability and Implementation:}
  \url{https://github.com/bio-ontology-research-group/walking-rdf-and-owl}
  \\
  \textbf{Contact:}
  \href{robert.hoehndorf@kaust.edu.sa}{robert.hoehndorf@kaust.edu.sa}

\section{Introduction}

The Semantic Web \cite{Berners-Lee2001}, a project with the stated
purpose of forming a consistent logical and meaningful web of data
using semantic technologies to make data machine-understandable and
processable, has been highly successful in biology and biomedicine
\cite{Katayama2014}. Many major bioinformatics databases now make
their data available as Linked Data in which both biological entities
and connections between them are identified through a unique
identifier (an Internationalized Resource Identifier or IRI) and the
connections between them are expressed through standardized relations
\cite{Wood2014, Smith2005}. Linked Data can enable interoperability
between multiple databases simply by reusing identifiers and utilizing
no-SQL query languages such as SPARQL \cite{sparql} that can perform
distributed queries over multiple databases. Some of the first major
efforts to make life science data available as Linked Data have been
the UniProt RDF initiative \cite{uniprot2015} and the Bio2RDF project
\cite{Belleau2008, Callahan2013}.  UniProt focuses on making data
within a single database, UniProt, available as Linked Data so that
information and identifiers can be reused in other databases, while
Bio2RDF has the aim to combine multiple databases and demonstrate the
potential of Linked Data in life sciences, in particular with regard
to provenance tracking, usability and interoperability. Now, major
databases, such as those provided by the European Bioinformatics
Institute (EBI) and the National Center for Biotechnology Information
(NCBI), are made available as Linked Data \cite{Jupp2014,
  Kim2016}. Additionally, community guidelines and principles for data
publishing such as the FAIR principles \cite{fair} require data to be
made available in a way that is amenable to interoperability through
linking and federation of queries.

A second major component of applications of the Semantic Web in the
life sciences has been the development and use of
ontologies. Ontologies are specificiations of a conceptualization of a
domain \cite{Gruber1995}, i.e., they formally and explicitly specify
some of the main regularities (classes of entities) that can be found
within a domain and their interconnections \cite{Hoehndorf2015role}.
Ontologies are now widely used in biological datasets for the
annotation and provision of metadata.  They are commonly represented
in formal languages with model theoretic semantics \cite{Grau2008,
  Horrocks2007} which makes them amenable to automated
reasoning. However, the large size of the ontologies and the
complexity of the languages and reasoning tasks involved have somewhat
limited ontologies' use in automated reasoning. In particular, there
is still a large disparity between the ontologies in biomedicine and
the databases that uses them for annotation. While inferences over the
ontologies, as part of ontology development and quality assurance
process, become increasingly common \cite{uberon, go2015}, they are
not always applied to infer new relations between biomedical data.


Recently, several machine learning methods have become available that
can be utilized to learn features from raw data
\cite{deeplearningreview}. Several of these methods can also be
applied to graph-structured data \cite{Yanardag2015,Perozzi2014}.
While most of these methods are developed for graphs without edge
labels (in contrast to Linked Data in which edge labels represent the
type of relation between entities), some methods have also been
extended to incorporate edge labels \cite{rdf2vec}. However, to be
applicable to biological data, a crucial aspect is the
interoperability between both the data layer (as expressed in Linked
Data formats) and the annotations of data items or semantic layer
(expressed through ontologies and the background knowledge they
provide).  This tight integration between data and knowledge, as
dominantly present in biological databases, benefits from automated
reasoning so that it becomes possible to consider inferred knowledge,
handle data consistency and identify incompatible conceptualizations.

We developed a method to leverage the semantic layer in knowledge
graphs such as the Semantic Web or Wikidata by combining
automated reasoning over ontologies and feature learning with neural
networks, to generate vector representations of nodes in these
graphs (node embeddings). We demonstrate that these representations
can be used to predict edges with biological meaning. In particular,
we demonstrate that our approach can predict disease genes, drug
targets, drug indications, gene functions and other associations with
high accuracy, in many cases matching or outperforming state of the
art methods.

Our results demonstrate how Linked Data and ontologies can be used to
form biological knowledge graphs in which heterogeneous biological
data and knowledge are combined within a formal framework, and that
these graphs can not only be used for data retrieval and search, but
provide a powerful means for data analysis and discovery of novel
biological knowledge.  

\section{Methods}
\subsection{Data Description}
In our experiments, we build a knowledge graph based on three
ontologies: the Gene Ontology (GO) \cite{Ashburner2000} downloaded on
18 July 2016, the Human Phenotype Ontology \cite{Koehler2014}
downloaded on 18 July 2016, and the Disease Ontology \cite{do}
downloaded on 19 August 2016. We also use the following biological
databases in our knowledge graph:
\begin{itemize}
\item Human GO annotations from SwissProt \cite{uniprot2015}, and
  phenotype annotations from the HPO databases \cite{Koehler2014}, 
  downloaded on 23 July 2016.  We include a total of 211,975 GO
  annotations and 153,575 phenotype annotations.
\item Human Proteins interactions from the STRING database
  \cite{Szklarczyk2011} downloaded on 18 July 2016. We filter proteins
  by their interactions confidence score and choose those above
  700. The total number of interactions in this dataset is 188,424.
\item Human chemical--protein interactions downloaded from the STITCH
  database \cite{Kuhn2012}, on 28 August 2016, filtered for confidence
  score of $700$. The total number of drug-target interactions present
  in the graph is 554,366.
\item Genes and disease associations from DisGeNET \cite{disgenet},
  downloaded on 28 August 2016, consisting of 236,259 associations.
\item Drug side effects and indications from SIDER \cite{Kuhn2010},
  downloaded on 15 August 2016. We include a total of 48,702
  drug--side effect pairs and 6,159 drug--indication pairs in our
  graph.
\item Diseases and their phenotypes from the HPO database
  \cite{Koehler2014} and text mining \cite{Hoehndorf2015srep}. We
  include a total of 84,508 phenotype annotations of diseases. 
\end{itemize}
We map all protein identifiers to Entrez gene identifiers and use
these to represent both genes and proteins. We use PubChem identifiers
to represent chemicals and we map UMLS identifiers associated with
diseases in DisGeNET and indications in SIDER to the Disease Ontology
using mappings provided by Disease Ontology. We further map UMLS
identifiers associated with side effects in SIDER to HPO identifiers
using mapping between UMLS and HPO \cite{Hoehndorf2013drugs}.

A knowledge graph is a graph-based representation of entities in the
world and their interrelations. Knowledge graphs are widely used to
facilitate and improve search, and they are increasingly being
developed and used through Semantic Web technologies such as the
Resource Description Framework (RDF) \cite{Candan2001}.  Here, we
focus on knowledge graphs centered around biological entities and
their interactions, ignoring all meta-data including labels or
provenance. The knowledge graphs we consider have two distinct types
of entities: biological entities, and classes from biomedical
ontologies that provide background knowledge about a domain.  The aim
of building a biological knowledge graph is to represent, within a
single formal structure, biological relations between entities, their
annotations with biological ontologies, and the background knowledge
in ontologies.

We make a clear distinction beween instances and classes. While there
is some debate about which kinds of biological entities should be
treated as instances and which as classes \cite{Smith2005}, similarly
to other Linked Data approaches \cite{uniprot2015}, we treat
biological entities such as types of proteins, diseases, or chemicals,
as instances in the knowledge graph.  In our case, classes from the
Disease Ontology are also treated as instances. On the level of
instances, we can integrate existing graph-based representations used
in biology and biomedicine, in particular biological networks such as
protein-protein interaction networks, genetic interaction networks,
metabolic interactions or pathways.

Ontology-based annotations are expressed by asserting a relation
between the instance (e.g., a disease or protein) and an instance of
the ontology class. For example, we express the information that the
protein {\em Foxp2} has the function {\em transcription factor
  binding} ({\tt GO:0003700}) by the two axioms
$hasFunction(foxp2, f_1)$ and $instanceOf(f_1, GO:0003700)$ where
$foxp2$ and $f_1$ are instances, $GO:0003700$ the class {\tt
  http://purl.obolibrary.org/obo/GO\_0003700} in GO, $hasFunction$ an
object property, and $instanceOf$ the {\tt rdf:type} property
specified in the OWL standard \cite{owl2-overview} as expressing an
instantiation relation. The instance $f_1$ can be expressed as an
anonymous instance (i.e., a blank node in the RDF representation) or
be assigned a unique new IRI. In our knowledge graph, we create a new
IRI (i.e., an IRI that does not occur anywhere else in the graph) for
each of these instances.

\subsection{Ontology-based classification}

Due to the large size of the knowledge graphs we process, we rely on
polynomial-time automated reasoning methods. OWL provides three
profiles \cite{owlprofiles} that facilitate polynomial time
inferences, and multiple RDF stores implement different subsets of OWL
to facilitate inferences and improve querying. For example, the
OWL-Horst subset \cite{owlhorst} is used by several RDF stores and is
useful in data management and querying. In biological and biomedical
ontologies, the OWL 2 EL profile is widely used to develop the large
ontologies that are in use in the domain, and has been found to be
useful and sufficient for a large number of tasks \cite{elvira,
  uberon, Schulzel2007}.

OWL 2 EL supports basic inferences over ontologies' class hierarchies
(including intersection, existential quantification and disjointness
between named classes), supports inferences over object properties
(transitivity, reflexivity, and object property composition), and can
infer the classification of instances. We make use of OWL 2 EL for
representing the knowledge graphs we generate and utilize the ELK
reasoner \cite{elk} for automated reasoning over them. In principle,
other profiles of OWL can also be used following a similar approach,
but may not be feasible due to the high computational complexity of
generating inferences \cite{dlhandbook}. OWL 2 EL supports the
following class descriptions, class and object property axioms (using
capital letters for classes, lower case letters for object properties,
and $x_1, x_2,...$ for instances):
\begin{itemize}
\item Class description: class intersection ($A \sqcap B$),
  existential quantification ($\exists r.A$), limited enumeration
  using a single instance ($\{x_1\}$)
\item Class axioms: subclass ($A\sqsubseteq B$), equivalent class
  ($A \equiv B$), disjointness ( $A\sqcap B \sqsubseteq \bot$)
\item Object property axioms: sub-property ($r \sqsubseteq s$),
  property chains ( $r \circ s \sqsubseteq q$), equivalent property (
  $r \equiv s$), transitive properties ($r \circ r \sqsubseteq r$),
  reflexive properties
\end{itemize}

We deductively close the knowledge graph with respect to the OWL 2 EL
profile, using an OWL 2 EL reasoner \cite{elk}. A knowledge graph
$\mathcal{KG}$ is deductively closed if and only if for all $\phi$
such that $\mathcal{KG} \models \phi$, $\phi \in \mathcal{KG}$.  In
general, the deductive closure of a knowledge is countably
infinite. Therefore, we only add inferences that can be represented
explicitly as edges between named individuals and classes in
$\mathcal{KG}$, i.e., between entities that are explicitly named in
$\mathcal{KG}$. In particular, for all instances
$x_i, x_j \in \mathcal{KG}$ and object properties (i.e., edge labels)
$r \in \mathcal{KG}$, if $\mathcal{KG} \models r(x_i, x_j)$, then
$r(x_i, x_j) \in \mathcal{KG}^{\models}$. Furthermore, for all named
classes $C \in \mathcal{KG}$ and instances $x \in \mathcal{KG}$, if
$\mathcal{KG} \models C(x)$, then $C(x) \in \mathcal{KG}^{\models}$.
Finally, we also infer relations between classes, in particular
subclass axioms, and add them to the inferred graph: for any class
$C, D \in \mathcal{KG}$, if $\mathcal{KG} \models C\sqsubseteq D$,
then $C \sqsubseteq D \in \mathcal{KG}^{\models}$.

We use the OWL API version 4 \cite{Horridge2007} to classify the input
knowledge graph and add all inferences obtained by using the ELK
reasoner as new edges to the knowledge graph to generate
$\mathcal{KG}^{\models}$.  We use this fully inferred graph as a basis
for generating the node embeddings through our method.

\subsection{Walking RDF and OWL}

To generate node embeddings, we use a modified version of the DeepWalk
algorithm \cite{Perozzi2014} in which we consider edge labels as part
of the walk. A random walk of length $n$ over a graph $G=(V,E)$ and
start vertex $v_0 \in V$ is an ordered sequence of vertices
$(v_0,...,v_n)$, $v_i \in V$, and each $v_i$ ($i>0$) is determined by
randomly selecting an adjacent node of $v_{i-1}$. As knowledge graphs
generated by our method additionally have edges of different types
(i.e., edge labels, $\mathcal{L}(E)$), we extend this notion to
edge-labeled random walks. An edge-labeled random walk of length $n$
over the graph $G=(V,E)$, edge labels $\mathcal{L}:E \mapsto L$ in the
label space $L$ (i.e., the set of object properties in the knowledge
graph underlying $G$), and start vertex $v_0 \in V$ is a sequence
$(v_0, l_1, v_1,..., l_{n}, v_n)$ such that $v_i \in V$, $l_i \in L$,
and, starting with $v_0$ and for all $v_i$ ($i<n$), a random outgoing
edge $e_{i+1}$ of $v_i$, ending in $v_{i+1}$ is chosen to generate
$l_{i+1}$ from $\mathcal{L}(e_{i+1})$ and $v_{i+1}$.
 
We implement this algorithm as an extension of the DeepWalk
\cite{Perozzi2014}. The algorithm takes a knowledge graph $G = (V,E)$
as input and generates a corpus $\mathcal{C}$ consisting of a set of
edge-labeled random walks, starting either from all vertices
$v \in V$, or all vertices $v \in U$ of a specified subset of
$U \subseteq V$. Parameters of the algorithm are the length of the
walks and the number of walks per node. Source code of the algorithm
and documentation are freely available at
\url{https://github.com/bio-ontology-research-group/walking-rdf-and-owl}.

\subsection{Learning Embeddings}
We use the corpus $\mathcal{C}$ of edge-labeled random walks as an
input for learning embeddings of each node. We follow the skip-gram
model \cite{skipgram} to generate these embeddings. Given a sequence
of words, $(w_1,...,w_N)$ in $\mathcal{C}$, a skip-gram model aims to
maximize the average log probability
\begin{equation}
\label{eqn:optimize}
\frac{1}{N} \sum_{n=1}^{N} \sum_{-c\leq j \leq c, j\not=
  0} \log p(w_{n+j}|w_n)
\end{equation}
in which $c$ represents a context or window size. To define
$p(w_{n+j}|w_n)$, we use negative sampling, following \cite{skipgram},
i.e., replacing $\log p(w_{O}|w_{I})$ above with a function to
discriminate target words ($w_O$) from a noise distribution $P_n(w)$
\cite{skipgram}, drawing $k$ words from $P_n(w)$:
\begin{equation}
\label{eqn:noise}
  \log \sigma(v_{w_O}^{' \intercal} v_{w_I}) + \sum_{i=1}^{k}
  \mathbb{E}_{w_i \sim P_n(w)} \left[\log \sigma (- v_{w_i}^{' \intercal}
    v_{w_I}) \right]
\end{equation}
The vector representation (embedding) of a word $s$ occurring in
corpus $C$ is the vector $v_s$ in Eqn. \ref{eqn:noise} derived by
maximizing Eqn. \ref{eqn:optimize}. The dimension of this vector is a
parameter of the method.

Since our corpus consists of often repeated edge labels (due to the
relatively small size of the label space $\mathcal{L}$), we further use
sub-sampling of frequent words \cite{skipgram} (which mainly represent
edge labels in the corpora we generate) to improve the quality of node
embeddings. We follow \cite{skipgram} and discard, during training,
each word $w_i$ (i.e., node or edge) with a probability
$P(w_i) = 1- \sqrt{\frac{t}{f(w_i)}}$ where $t$ is a threshold
parameter.

It is obvious from this formulation that the parameters for learning
the representation of nodes in a knowledge graph include the number of
walks to perform for each vertex, the length of each individual walk,
a subset $U$ of vertices from which to start walks, the size of the
vector representations learned by the skip-gram model, the window or
context size employed in the skip-gram model, the parameter $t$ used
to sub-sample frequent words (we use $t=10^{-3}$ for all our
experiments), and the number of words to draw from the noise
distribution (we fix this parameter to $5$ in our experiments).  There
are several additional parameters for training a skip-gram model,
including learning rate and certain processing steps on the corpus,
for which we chose default values in the gensim
(\url{https://radimrehurek.com/gensim/}) skip-gram implementation.

\subsection{Prediction}
The embeddings can be used as features in machine learning tasks that
should encode for the local neighborhood of each node, thereby
encoding for the (local) information contained in a knowledge graph
about a certain vertex. We apply these features to the task of edge
prediction, in which we aim to estimate the probability that an edge
with label $l$ exists between vertices $v_1$ and $v_2$ given their
vector representation, $\mathbb{v}(v_1)$ and $\mathbb{v}(v_2)$:
$p((v_1,v_2,l) \in E | \langle \mathbb{v}(v_1), \mathbb{v}(v_2)
\rangle )$.
We use the logistic regression classifier implemented in the sklearn
library \cite{scikit-learn} to train logistic regression models.

We build separate binary prediction models for each edge type in the
knowledge graphs. For model building and testing, we employ 5-fold
cross-validation. Each cross-validation fold is built by randomly
removing 20\% of edges of a particular type in the knowledge graph,
then applying deductive inference, corpus generation through
edge-labeled random walks, learning of vector representations of
nodes, and building of a binary logistic regression model. A model for
edges with label $l$ is trained using as positive instances all pairs
of vertices for which an edge with label $l$ exists in the modified
knowledge graph (in which 20\% of edges with label $l$ have previously
been removed), and using as negatives a random subset of all pairs of
vertices $(v_{r_1}, v_{r_2})$ such that $v_{r_1}$ is of the same type
(i.e., an instance of the same class in the knowledge graph) as all
sources of edges with label $l$, and $v_{r_2}$ is of the same type as
all targets of edges with label $l$. For example, if edges with label
$l$ are all between instances of {\em Drug} and {\em Disease} in a
knowledge graph, then we sub-sample negative instances among all pairs
of instances of {\em Drug} and {\em Disease} for which no edge exists
in the original knowledge graph. The constraint of choosing negatives
from the same general types of entities is necessary because instances
of different types will be clearly separable within the embeddings,
and evaluation using those would therefore bias the results.  We
randomly generate a set of negative samples with the same cardinality
as the set of positive samples, both for model training and
prediction.

The embeddings can also be used for findings similar nodes using a
measure of similarity. We use cosine similarity to compute the
similarity between two vectors: $sim(v_1, v_2) = \frac{v_1 \cdot v_2}{\lVert A \rVert \lVert B
  \rVert}$

\subsection{Parameter optimization}

Using the performance on the final prediction model, we perform
parameter optimization through a limited grid search. We only optimize
embedding size, number of walks, walk length and context size for the
skip-gram model through a grid search since an exhaustive optimization
would be too computationally expensive. Furthermore, we only use a
single edge type to test how results change with each choice of
parameter, due to computational constraints. We tested the following 625 parameters
: embedding sizes of $32$, $64$, $128$, $256$, and $512$, number of walks $50$, $100$,
$200$, $300$, and $500$, walk length $5$, $10$, $15$, $20$, and $30$,
and skip-gram context sizes $5$, $10$, $15$, $20$, and $30$. We found
the best performing parameters to be $512$ for the embedding size,
$100$ for the number of walks, $20$ for the walks length and $10$ for
the skip-gram context size, and we fix these parameters throughout our
experiments.

\section{Results}
\subsection{Neuro-symbolic feature learning using Semantic Web
  technologies}

We build a knowledge graph using Semantic Web technologies centered on
human biomedical data. The graph incorporates several biological and
biomedical datasets and is split in two layers, instances and classes.
On the level of instances in the knowledge graph, we combine
protein-protein interactions (PPIs) \cite{Szklarczyk2011}, chemicals
(drugs) and their protein targets \cite{Kuhn2012}, drugs and their
indications \cite{Kuhn2010}, and genes and the diseases they are
involved in \cite{disgenet}.  On the level of classes, we include the
Human Phenotype Ontology \cite{Koehler2014},
and the Gene Ontology \cite{Ashburner2000}, and we include annotations
of diseases and their phenotypes \cite{Koehler2014,
  Hoehndorf2015srep}, genes and their phenotypes \cite{Koehler2014},
and human protein functions and subcellular locations
\cite{uniprot2015}.  The knowledge graph, including the data,
ontologies and our formal representation of ontology-based
annotations, consists of 7,855,737 triples.  We use the Elk reasoner
\cite{elk} to deductively close this graph, and through the
application of ontology-based inference, we further infer 5,664,387
new triples and add them to the knowledge graph. 


We utilize this knowledge graph as the input to our algorithm that can
learn representations of nodes. These representations represent the
neighborhood of a node as well as the kind of relations that exist to
the neighboring nodes. To learn these representations, we perform
random walks from each node in the knowledge graph repeatedly, use the
resulting walks as sentences within a corpus, and apply the Word2Vec
skip-gram model \cite{skipgram} to learn embeddings for each node.

We use the fully inferred, deductively closed knowledge graph to
perform the random walks.  Performing random walks on the deductively
closed graph has the advantage that not only asserted axioms will be
taken into consideration, but representations can also include
inferred knowledge that is not present explicitly in the graph.  
For example, for an assertion that a gene $g$ has a function $F$
(where $F$ is a class in the GO), all superclasses of $F$ in GO will
be added as annotations to $g$; sub-properties (such as
$binds \subseteq \mbox{interacts-with}$) asserted in an ontology or
database will be resolved; transitive, reflexive object properties and
property chains
resolved and the inferred edges added.

\begin{figure*}[ht!]
  \centering
  \includegraphics[width=.76\textwidth, bb=14 14 4030 1930]{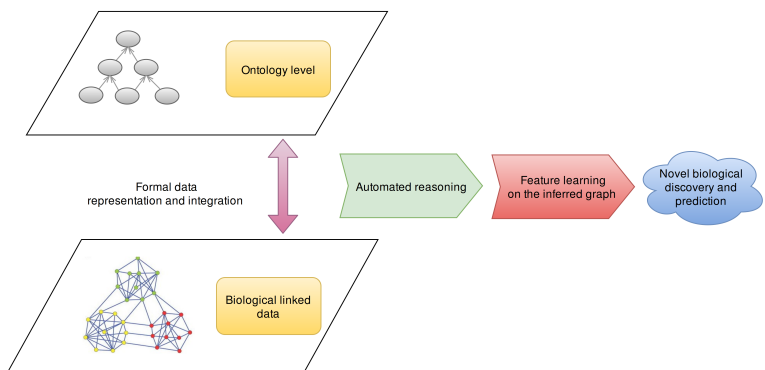}
  \caption{\label{fig:overview}Overview over the main steps in our
    workflow. We first build biological knowledge graphs by integrating
    Linked Data, biomedical ontologies, and ontology-based annotations in
    a single, two-layered graph, then deductively close the graph using
    automated reasoning and apply feature learning on the inferred graph
    to take into account both explicitly represented data and inferred
    information.}
\end{figure*}

We automated these steps (ontology-based classification, repeated
random walk, generation of embeddings) in an algorithm that combines
the steps relying on symbolic inference and the learning of embeddings
using a neural network. The input of the algorithm is a knowledge
graph and the parameters needed for the algorithm such as the length
and number of walks and size of the resulting embeddings, the output
is an embedding (of a specified size) for each node in the knowledge
graph. Figure \ref{fig:overview} illustrates our basic workflow.

\subsection{Edge Prediction}
The resulting embeddings can be used in standard machine learning
classifiers. We demonstrate these uses in two settings. First, we
remove edges from the knowledge graph, regenerate the embeddings using
the reduced graph, and train a logistic regression classifier to
predict whether or not an edge exists between two nodes, given the
embeddings for two nodes as input. This kind of application is
intended to demonstrate how associations between two potentially
different types of entities, such as a gene and disease, can be
identified. We perform these experiments in 5-fold cross-validation
setting for every edge type in our graph except edges that exist only
between ontology classes. Table \ref{tab:performance} summarizes the
results.

\begin{table}[ht!]
\caption{\label{tab:performance}Performance results for edge
  prediction in a biological knowledge graph. Edge types marked with
  an asterix are between instances and instances of ontology classes.} 
\centering
\begin{tabular}{@{}lcccc@{}}\toprule 
\multirow{2}{*}{Edge Type} 
&\multicolumn{2}{c}{Without reasoning}&\multicolumn{2}{c}{With reasoning}\\
           & F-measure & AUC & F-measure & AUC \\
\midrule
has target & 0.94 & 0.97 & 0.94 & 0.98 \\
has disease annotation & 0.89 & 0.95 & 0.89 & 0.95 \\
has side-effect$^*$ & 0.86 & 0.93 & 0.87 & 0.94 \\
has interaction & 0.82 & 0.88 & 0.82 & 0.88\\
has function$^*$ & 0.85 & 0.95 & 0.83 & 0.91 \\
has gene phenotype$^*$  & 0.84 & 0.91 & 0.82 & 0.90  \\
has indication & 0.72 & 0.79 & 0.76 & 0.83 \\
has disease phenotype$^*$  & 0.72 & 0.78 & 0.70 & 0.77 \\
\bottomrule
\end{tabular}
\end{table}

We find that the performance of the prediction differs significantly
by edge type, but some types of relations can be predicted with high
F-measure. Furthermore, using the knowledge graph with reasoning
improves the performance slightly when predicting edges between
instances and mostly results in decreased performance when aiming to
predict edges between instances and and instance of an ontology
class. We achieve overall highest performance on predicting {\em has
  target} edges with an F-measure of $0.94$ and ROCAUC of $0.98$, and
lowest overall performance on associations between diseases and their
phenotypes ({\em has disease phenotype}, ROCAUC $0.77$). While our aim
here is not to propose a novel method of predicting drug targets,
protein functions or phenotypes, state of the art approaches that
incorporate multiple types of information and use graph inference for
predicting drug--target edges achieve ROCAUC of up to $0.93$
\cite{drugtarget1, Wang2014}, albeit using a different set of positive
and negative drug--target pairs. Similarly, some of the edges, such as
{\tt has function} or {\tt has phenotype}, have to be predicted in a
hierarchical output space (i.e., an ontology such as the Gene Ontology
\cite{Ashburner2000} and the Human Phenotype Ontology
\cite{Koehler2014}) and need to satisfy additional consistency
constraints (due to formal dependencies between the labels), which may
overall result in lower performance when applied to these tasks
\cite{cafa, Sokolov2013}. While the evaluation results demonstrate
that the embeddings of vertices learned through our approach contain
sufficient information about the node to be useful in predictive
models, more specific knowledge graphs, containing different types of
information, need to be built for specific applications.

\subsection{Drug repurposing on biological knowledge graphs}

As second use case, we also test how well the node embeddings can be
used to predict novel relations, i.e., relations that are not
explicitly represented in the knowledge graph. Such an evaluation
can provide information about how well the embeddings our algorithm
generates can be reused in novel applications or as part of larger
predictive systems for hypothesis generation \cite{Gottlieb2011}.

We aim to test how much information about shared mode of action is
encoded in the embeddings of drug nodes generated by our method, and
how the performance of our approach compares to related efforts.
Using side-effect similarity alone, it is possible to identify pairs
of drugs that share protein targets and indications
\cite{Campillos2008, Tatonetti2012}, thereby demonstrating that side
effects provide some information about drugs' modes of action
\cite{Campillos2008}.  We train a logistic regression classifier to
predict whether a pair of drugs (represented by the embeddings we
generate) share an indication or target. To make our input data
comparable to studies that compare only drugs' side effects, and to
avoid bias introduced by encoding targets and indications in the
knowledge graph, we remove all {\em has indication} and {\em
  has target} edges from our graph and further retain only drugs
contained in the SIDER database \cite{Kuhn2010}. We then train a
logistic regression classifier to determine whether a pair of drugs
shares an indication, a target, or both, using 80\% of the drug pairs
as training and keeping 20\% as testing.

\begin{figure}[ht!]
  \centering
  \includegraphics[width=.5\textwidth]{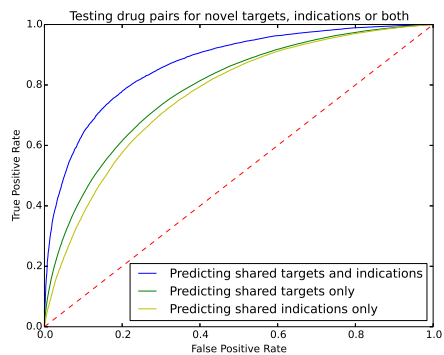}
  \caption{\label{fig:rocauc}ROCAUC test scores of SIDER drug pairs
    the for predicting novel indications or targets or both.}
\end{figure}

Figure \ref{fig:rocauc} shows the resulting performance. We can
achieve up to $0.87$ ROCAUC for predicting pairs of drugs that share
both an indication and a target, $0.79$ ROCAUC for drugs that share
targets, and $0.77$ ROCAUC for drugs that share indications.  In
comparison, ranking drug pairs by their side effect similarity alone
can achieve a ROCAUC of up to $0.75$ for drugs sharing targets and
$0.83$ for drugs sharing indications \cite{Tatonetti2012}.  Our
results demonstrate that our method generates embeddings that encode
for the explicit information in a knowledge graph, is capable of
utilizing this for prediction and achieve comparable results to other
approaches. Moreover, after removing {\em has target} and {\em has
  indication} edges, drugs are not directly linked to protein-protein
interactions, protein functions or disease phenotypes. Nevertheless,
the embeddings generated for drugs based on the corpus generated by
random walks can encode some of this information, for example by
linking both genes and drugs to similar phenotypes (and thereby
providing information about potential drug targets), linking diseases
and drugs to similar phenotypes (and thereby providing information
about potential indications), as well as more complex interactions.

Instead of using a classifier, similarity between the embeddings can
also be exploited to identify biological relations. Using the full
knowledge graph, we further tested whether drug-drug similarity can be
used to identify drugs that fall in the same indication group. We use
cosine similarity to determine how similar two drugs are and evaluate
whether drugs that share the same top-level Anatomical Therapeutic
Chemical Classification System (ATC) code are more similar than drugs
that do not share codes. We find drugs in the same ATC top-level
category are significantly ($p < 3 \cdot 10^{-4}$ , Mann-Whitney U
test) more similar than drugs that do not fall in the same ATC
top-level category.

\section{Discussion}
%
A limitation of our approach is the reliance on qualitative data and
ignoring quantitative information. Many of the edges in biological
networks have associated weights that represent either the strength of
a biological relation or the certainty of the existence of a
relation. For example, networks which are based primarily on
correlation between quantitative, experimentally measured variables
use correlation coefficients to quantify the strength of a biological
association, while network databases such as STRING
\cite{Szklarczyk2011} include data from multiple databases and
associate a confidence value for the existence of an edge. While
random walk algorithms can, in principle, be extended to account for
edge weights \cite{Coppersmith1993}, the different semantics of the
edge weights will require different treatments by the algorithm. In
particular, incorporating edge weights may need to distinguish between
weights that represent confidence in the existence of a relation (as
in confidence scores from a prediction method) and weights that
quantify the biological strength of the relation (as in correlation
coefficients).

Despite the large success of machine learning methods in the past
years \cite{deeplearningreview}, they have not yet widely been applied
to symbolically represented biological knowledge. Symbolic
representations in biology, based on Linked Data and ontologies, are
relying on formal languages such as OWL and RDF, and utilize symbolic
inference. The kind of inferences performed on this knowledge is
either formally specified in the knowledge representation language
\cite{dlhandbook} or arised from hand-crafted inference rules that are
applicable within a particular database, application, or query
\cite{Callahan2013, uniprot2015}. Here, we use knowledge graphs built
using the semantics of OWL and data is represented as instances of OWL
classes, but our approach of building knowledge graphs can be replaced
with, or amended by, the use of explicit inference rules. In this
case, instead of applying an OWL reasoner to infer edges with respect
to the OWL semantics, rules can be used to infer edges and deductively
close the knowledge graph with respect to a set of inference rules.

A key difference between the knowledge graphs we use in our approach
and knowledge graphs widely used in biological databases is the strong
focus on representing biological entities and their relations in
contrast to representing the (non-biological) meta-data about these
entities and their associations, such as provenance \cite{provo} and
authorship. While inclusion of such metadata in knowledge graphs is
required for retrieval and to ensure data quality \cite{fair}, our
method relies on the use of data models that make it possible to
separate the biological content of a knowledge graph from the
metadata. 

We demonstrate that knowledge graphs based on Semantic Web standards
and technologies can not only be used to store and query biological
information, but also have the capability to model and represent
biological phenomena, such as the totality of known protein-protein
interactions within a cell. The key advantage of choosing knowledge
graphs over other representations is the inherent focus on
representing heterogeneous information in contrast to single types of
relations, the possibility to continuously add information, the use of
inference rules, and the use of World Wide Web standards. Our method
allows all these advantages to be utilized for data analysis and to
build predictive models, and may encourage database curators and
biologists to increasingly rely on knowledge graphs to represent the
biological phenomena of their interest.





\section*{Acknowledgements}

A prototype of the feature learning algorithm was implemented at the
NBDC/DBCLS BioHackathon 2016 in Tsuruoka.

\section{Funding}
This work was supported by funding from King Abdullah University of
Science and Technology (KAUST).


\bibliographystyle{plainnat}


\end{document}